\documentclass[osajnl,twocolumn,showpacs,superscriptaddress,10pt]{revtex4-1} 
\usepackage{amsmath,amssymb,graphicx}
\usepackage{geometry} 
\begin{document}

\title{Bayesian-based aberration correction and numerical diffraction for improved lensfree on-chip microscopy of biological specimens}

\author{Alexander Wong}\email{Corresponding author: a28wong@uwaterloo.ca}
\author{Farnoud Kazemzadeh}
\affiliation{Department of Systems Design Engineering, University of Waterloo, 200 University Ave. West, Waterloo, Ontario, N2L 3G1}
\author{Chao Jin}
\affiliation{Department of Civil and Environmental Engineering, University of Waterloo, 200 University Ave. West, Waterloo, Ontario, N2L 3G1}
\author{Xiao Yu Wang}
\affiliation{Department of Systems Design Engineering, University of Waterloo, 200 University Ave. West, Waterloo, Ontario, N2L 3G1}

\begin{abstract}Lensfree on-chip microscopy is an emerging imaging technique that can be used to visualize and study biological specimens without the need for imaging lens systems.  Important issues that can limit the performance of lensfree on-chip microscopy include interferometric aberrations, acquisition noise, and image reconstruction artifacts.  In this study, we introduce a Bayesian-based method for performing aberration correction and numerical diffraction that accounts for all three of these issues to improve the effective numerical aperture (NA) and signal-to-noise ratio (SNR) of the reconstructed microscopic image.  The proposed method was experimentally validated using the USAF resolution target as well as real waterborne \emph{Anabaena flos-aquae} samples, demonstrating improvements in NA by $\sim$25\% over the standard method, and improvements in SNR of 2.8 dB and 8.2 dB in the reconstructed image when compared to the reconstructed images produced using the standard method and a maximum likelihood estimation method, respectively.
\end{abstract}


\maketitle 

Lensfree on-chip microscopy~\cite{lensfree1,lensfree2,lensfree3,greenbaum,bio1} is an emerging imaging microscopy technique capable of non-contact, high resolution imaging of specimens at the sub-micron scales.  In lensfree on-chip microscopy, a coherent or partially coherent light source is utilized to illuminate a specimen. The light-matter interaction results in an interference pattern which is then observed and digitally acquired on an array detector as a hologram. Using numerical diffraction, microscopic images can be reconstructed from the holograms.

There are several benefits to lensfree on-chip microscopy.  Since the specimen is placed very close to the image sensor array (e.g., $<$ 5 mm), it allows for highly compact and field-portable microscopy systems.  Furthermore, contrary to conventional lens-based microscopy techniques, the field-of-view (FOV) is equal to the active area of the image sensor array and as such allows for large FOVs while maintaining high spatial resolution that increase with decreasing pixel pitch in new sensor arrays~\cite{greenbaum}.  Additionally, lensfree on-chip microscopy facilitates naturally for quantitative phase contrast microscopy without the need to distort the bright field image like more complex methods such as differential interference contrast (DIC) microscopy.  Therefore, given all these benefits as well as its ability to visualize and quantify optical properties of transparent and semi-transparent specimens in a non-invasive manner, lensfree on-chip microscopy has begun to gain significant interest for the visualization and study of biological specimens such as sperm~\cite{bio1}, waterborne algae~\cite{bio4}, waterborne parasites~\cite{bio5}, and premalignant/malignant cells~\cite{bio2,bio3}.

There are several important issues that can limit the overall performance of lensfree on-chip microscopy systems.  First, the NA of a lensfree on-chip microscopy system is limited by the pixel size of the sensor array used. Second, system factors such as the pixel responsivity of the image sensor array~\cite{greenbaum} can result in interferometric aberrations that appear as distortions and/or defocusing, hence adversely affecting the effective NA of the reconstructed image.  Third, the quality of the reconstructed image is highly sensitive to the SNR of the recorded hologram, making it difficult to achieve high image reconstruction performance under low SNR imaging conditions.  Fourth, due to the nature of the recorded hologram and the numerical diffraction process, reconstruction artifacts such as ringing artifacts are often introduced, thus also reducing the effective SNR of the reconstructed image.  As such, methods for dealing with these important issues are highly desired to improve the effective NA and SNR of the reconstructed image for lensfree on-chip microscopy systems.

Much of existing literature in lensfree on-chip microscopy has focused on dealing with the first issue of pixel size limitations through the use of superresolution techniques~\cite{bio3,greenbaum,super1,super2,super3,super4}.  In such approaches, a stack of lower resolution holograms of the same specimen, acquired at different subpixel offsets thus each containing unique information about the specimen, are numerically combined to form a higher resolution hologram.  While such superresolution techniques can significantly improve the effective NA, they require more complex components in the optical system (e.g., mechanical micro-stage~\cite{super3}) to achieve subpixel offsets.  Fewer existing literature in lensfree on-chip microscopy have dealt with the second issue of interferometric aberrations.  Of particular interest is the seminal work by~\cite{greenbaum}, where these aberrations were characterized down to the pixel responsivity of the sensor array, and a maximum likelihood (ML) estimation approach~\cite{Richardson} was used to perform aberration correction on the recorded hologram based on the characterization.  While this resulted in significant gains in effective NA, it also resulted in significant undesirable reconstruction artifacts.

\begin{figure}[]
	\centering
    \includegraphics[width=0.4\linewidth]{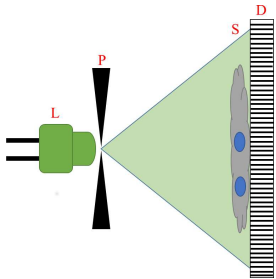}
        \vspace{-0.15in}
	\caption{Experimental lensfree on-chip microscopy setup. L is the light source, P is the aperture, a 75 $\mu$m pinhole, S is the sample, and D is the detector.}
	\label{fig1}
\vspace{-0.25in}
\end{figure}

To deal with the issue of hologram noise, methods such as~\cite{greenbaum} taking the Poisson characteristics of photon noise were taken into account in the ML estimation approach when performing aberration correction; however, the noise persists in the reconstructed image thus leading to reduced SNR.  Furthermore, for the case of lensless on-chip microscopy of biological specimens, the variance of parameter estimates are high for the ML approach, leading to results that are highly dominated by variations in data.  To perform further noise removal and compensation, post-processing methods have been proposed without taking the inherent noise statistics into account~\cite{denoise,denoise2,denoise3,denoise4}.  Furthermore, the issue associated with reconstruction artifacts such as ringing artifacts due to the nature of the numerical diffraction process are not well explored.  In addition, each of these issues are primarily dealt with one at a time in existing literature, which can lead to suboptimal results as the issues are not taken into consideration concurrently.

In this study, we investigate and attempt to mitigate the last three aforementioned issues (interferometric aberration, acquisition noise, and reconstruction artifacts) by introducing a Bayesian-based method for performing aberration correction and numerical diffraction that takes into account all three issues in unison.  By incorporating prior models related to interferometric aberrations, acquisition noise, and reconstruction artifacts, one can compensate for these issues within a unified framework to improve the effective NA and SNR of the reconstructed image when compared to addressing these issues independently.  Furthermore, the proposed Bayesian-based method employs a Maximum a Posteriori (MAP) strategy, which provides improvements over an ML approach by adding bias which leads to variance reduction. This is especially beneficial for the application of lensless on-chip microscopy of biological specimens and acts as a key novelty of the proposed work.

In the proposed method, we model the desired lensfree on-chip microscopy image ($f$) (at $z$ with wavelength $\lambda$), and the measured hologram ($g$) as probability distributions.  In the proposed Bayesian-based method, the goal is to determine the most probable desired image $\hat{f}$ given the measured hologram $g$, based on prior knowledge related to $f$, knowledge of the transfer function associated with aberrations ($H_a$), as well as the numerical diffraction transfer function ($H_{d,z,\lambda}$) (the Fresnel transform, Huygens convolution, and angular spectrum methods for numerical diffraction can all be expressed as a convolution).  This can be formulated as the following MAP problem:
\begin{equation}
	\hat{f} = {\rm argmax}_{f}~p\left(f | g \right),
\end{equation}
\noindent where $p\left( f | g \right)$ is the conditional probability of $f$ given $g$.  By solving this problem, we effectively perform both numerical diffraction and aberration correction to obtain a corrected, reconstructed image.  This problem can be equivalently formulated as:
\begin{equation}
	\hat{f} = {\rm argmax}_{f}~p\left(g | f\right)p(f),
\label{MAP}
\end{equation}
\noindent where $p(g | f)$ is the likelihood and $p(f)$ is the prior.

\begin{figure}[]
	\centering
    \includegraphics[width=1\linewidth]{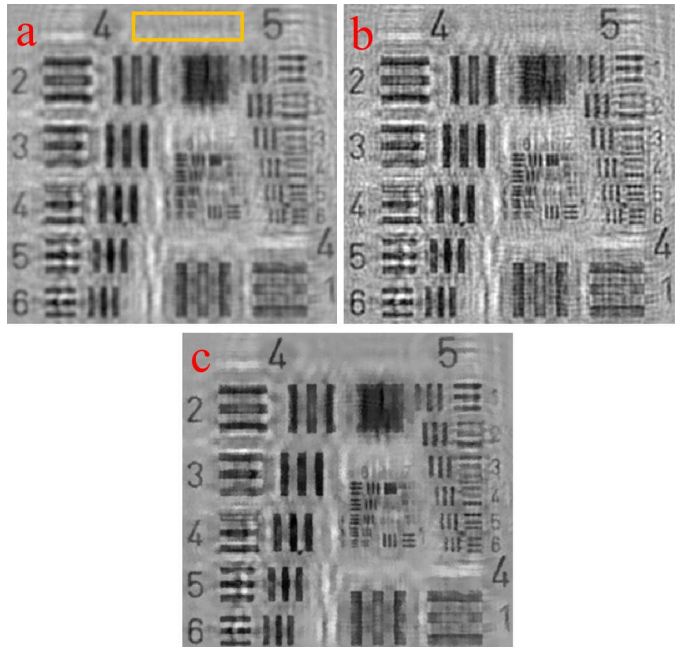}
    \vspace{-0.3in}
	\caption{A region of the USAF resolution target results. Reconstructed lensfree images using (a) standard method, (b) ML estimation~\cite{greenbaum}, and (c) the proposed method.  The SNR was assessed in a region shown by the orange box, and the DI~\cite{bianco} was evaluated in the orange box (to assess noise level) as well as the whole image (to assess contrast improvement).}
	\label{fig2}
\vspace{-0.2in}
\end{figure}

Let $S$ be a set of pixel locations in the image sensor array and $s \in S$ be a specific pixel location in the sensor array.  In the case of lensfree on-chip microscopy, acquisition noise is primarily related to quantum photon emission.  As such, to account for the statistics of this noise in the proposed method, along with aberrations (characterized by $H_a$) and numerical diffraction (characterized by $H_{d,z,\lambda}$), we incorporate the following likelihood $p(g|f)$:
\begin{equation}
	p\left(g | f\right) = \prod_{s \in S} \frac{\left(\mathfrak{F^{-1}}\left\{\frac{H_a}{H_{d,z,\lambda}}\mathfrak{F}\left\{f_s\right\}\right\}\right)^{{g_s}}e^{-(\mathfrak{F^{-1}}\left\{\frac{H_a}{H_{d,z,\lambda}}\mathfrak{F}\left\{f_s\right\}\right\})}}{{g_s}!}
\label{likelihood}
\end{equation}
\noindent where $\mathfrak{F}$ and $\mathfrak{F^{-1}}$ denotes the forward and inverse Fourier transform, respectively.  To compensate for the presence of reconstruction artifacts such as ringing artifacts due to the nature of lensfree on-chip microscopy, we explicitly enforce a prior model $p(f)$ where we model $f$ as a nonstationary process with a nonstationary expectation ${E}(f_s)$ and a variance $\tau^2$:
\begin{equation}
	p\left(f\right) = \prod_{s \in S} e^{-\frac{\left(f_s-{E}(f_s)\right)^2}{2 \tau^2}}.
\label{prior}
\end{equation}
\begin{figure}[!t]
	\centering
    \includegraphics[width=1\linewidth]{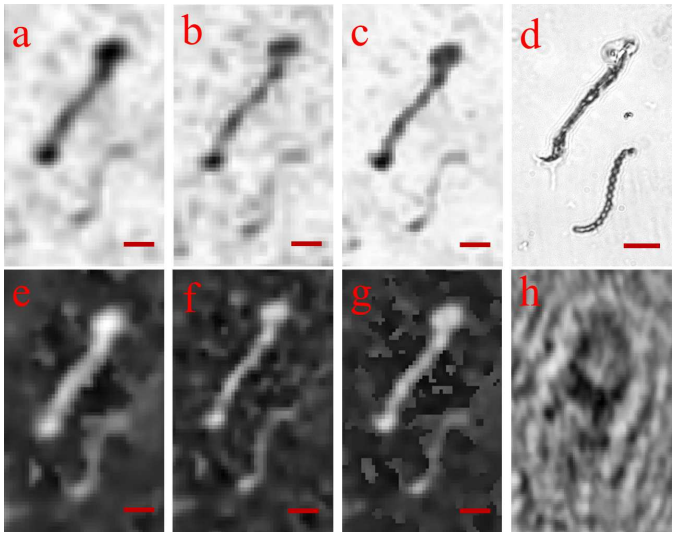}
            \vspace{-0.2in}
	\caption{Zoomed-in regions of the \emph{Anabaena flos-aquae} sample. (h) acquired hologram; (d) reference 40X microscopy image; reconstructed lensfree amplitude images using (a) standard method, (b) ML estimation~\cite{greenbaum}, (c) and the proposed method; reconstructed lensfree phase images using (e) standard method, (f) ML estimation, (g) and the proposed method. Contrast adjusted for better visualization. The scale bar denotes 20 $\mu$m.}
	\label{fig3}
\vspace{-0.2in}
\end{figure}

In the proposed method, the nonstationary expectation ${E}(f_s)$ is estimated via the nonparametric expectation estimation method introduced in~\cite{Wong} as it was shown to provide reliable estimates even under low SNR scenarios.  As such, the issues associated with interferometric aberrations, acquisition noise, and reconstruction artifacts are taken into account in unison in the proposed method via the likelihood $p(g|f)$ and prior model $p(f)$ presented in Eq.~\ref{likelihood} and Eq.~\ref{prior} to provide an improved estimate of $f$.  To obtain a corrected, reconstructed image $\hat{f}$, the MAP problem posed in Eq.~\ref{MAP} and is solved using the iterative optimization method described in~\cite{Wong}.

To demonstrate the efficacy of the proposed method for improving effective NA and SNR of the reconstructed image for lensfree on-chip microscopy, the method was validated using both the USAF resolution target and waterborne \emph{Anabaena flos-aquae} samples.  For this study, a lensfree on-chip microscope was constructed (see Fig.~\ref{fig1} for experimental configuration).  The partially-coherence light source used was a $\lambda=525$ nm, 7 mW light emitting diode (Thorlabs LED528EHP), and the interferometric hologram was observed and digitally acquired using a 1600$\times$1200 pixel CMOS sensor array (IDS-UI-1250LE-M-GL) with a pixel pitch of 4.5 $\mu$m.  The FOV of the microscope is determined by the active sensor size and is $\sim$ 39 mm$^2$ .  The lensfree on-chip microscope was characterized based on image acquisitions of the USAF resolution target to estimate the aberration transfer function $H_a$ using a least-squares optimization approach relative to an aligned ground-truth USAF resolution target image.  Given the recorded holograms from the lensfree on-chip microscope, the aberration transfer function $H_a$, and the Huygens numerical diffraction transfer function $H_{d,z,\lambda}$, the proposed method was run for 35 iterations to produce the final microscopic image.  For comparison purposes, the maximum likelihood (ML) estimation method described in~\cite{greenbaum} was also tested, with the number of iterations also set to 35 iterations and Hyugens numerical diffraction used for experimental consistency.  As a baseline reference, direct Hyugens numerical diffraction of the recorded hologram is also performed (we will refer to this as the standard method).

Fig.~\ref{fig2} shows a zoomed-in region from the lensfree amplitude images of the USAF resolution target using the standard method, ML estimation~\cite{greenbaum}, and the proposed method.  It can be observed that the horizontal and vertical lines of group 6 element 2 are resolved using the standard method, while the horizontal lines of group 6 element 6 and vertical lines of group 6 element 5 are resolved using the proposed method. These elements are considered to be resolved if the the width of the lines and separation between the lines are spanned by at least two pixels in the reconstructed images. The per-pixel resolution of the lensfree microscope is then determined by the separation distance between the lines of the USAF target divided by the number of pixels spanning the said distance.  The proposed method was able to achieve a gain of $\sim$ 25\% in NA over the reconstructed lensfree amplitude image using the standard method.

The SNR for all three lensfree amplitude images was assessed, with the SNR of the reconstructed lensfree amplitude image using standard method at 27.2 dB, the SNR of the reconstructed lensfree amplitude image using the ML estimation method at 21.8 dB, and the SNR of the reconstructed lensfree amplitude image using the proposed method at 30.0 dB.  Based on these results, it can be observed that the proposed method was able to achieve a gain of 2.8 dB over the standard method and 8.2 dB over the ML estimation method.

Furthermore, the Dispersion Index (DI)~\cite{bianco} was assessed for a homogeneous region (see Fig.~\ref{fig2}(a)) to assess noise level (lower DI here indicates lower noise levels), and for the whole image to assess contrast enhancement (higher DI here indicates better contrast enhancement). The DI achieved was 0.04/0.27 (homogeneous/whole-image) using the standard method, 0.08/0.30 using the ML estimation method, and 0.03/0.29 using the proposed method.  Based on these results, it can be observed that the proposed method achieved a lower DI in the homogeneous region that the standard and ML methods, which indicates that the proposed method achieved the lowest noise level. Furthermore, the proposed method achieved a higher whole-image DI than the standard method and comparable whole-image DI to the ML method, which indicates that the proposed method achieved improved contrast compared to the standard method and similar contrast as the ML method.

To validate the proposed method using real biological samples, cells of a laboratory pure culture of the cyanobacteria \emph{Anabaena flos-aquae} which is a neurotoxin producer common in surface water was selected as target for image acquisition (shown in Fig.~\ref{fig3}).  After two weeks of continuous culturing at 23$^o$ in a nutrient rich medium under illumination using a 20 W aquarium bulbs as light source, the selected \emph{Anabaena flos-aquae} reached a steady state with a cell count of $\sim$10$^6$ cells/mL.  An aliquot of 20 $\mu$L of the prepared sample was mounted on a pre-cleaned quartz slide for use in this study. Before image acquisition, the sample was placed in a biosafety cabinet for 30 minutes to ensure sedimentation and stabilization. Aiming to provide a visual reference, an inverted fluorescence microscope (Eclipse Ti, Nikon, Canada), at a total magnification of 40X with a resolution of 0.26 $\mu$m was used to obtain a bright field microscopy image of the specimen.

Fig.~\ref{fig3} demonstrates the detailed comparison between all mentioned methods, amplitude (a-c) and phase (e-g),  with a 40X microscopy image (d) and raw recorded hologram (h) shown for reference.   In the raw recorded hologram Fig.~\ref{fig3}(h), it can be observed that, due to the use of a realistic cultured sample where there may be contamination during media culture preparation, the interference patterns are highly complex, making it difficult to use post-processing methods on the hologram directly to compensate for noise without affecting the patterns.  Looking at the reconstructed lensfree images using standard method, Fig.~\ref{fig3}(a and e), and the reconstructed lensfree images using ML estimation Fig.~\ref{fig3}(b and f), the coupled \emph{Anabaena} filaments on the right side image were difficult to see due to interferometric aberrations (in the case of the standard method that does not account for such aberrations) and low SNR (in both cases), which could lead to underestimation of the number and volume of microorganisms.  This type of specific underestimation due to numerical error could 'artificially' increase the threshold and trigger corresponding treatment actions from decision makers during water treatment (e.g. higher dosage of disinfectant or extra engineered treatment steps) or potentially increase the human health risk due to exposure to neurotoxin released by \emph{Anabaena} in water used for human consumption.

In contrast, the reconstructed lensfree images using the proposed method, Fig.~\ref{fig3}(c and g), allows for a more reliable differentiation between the \emph{Anabaena} and the background due to improved SNR. Furthermore, additional detailed morphological features can be distinguished such as the hooked tail of the species.  In applied microbiology, this type of detailed information is very useful for accurate detection, enumeration and identification of each specific genus or species of microorganism present in environmental samples.  Note that while the FOV in the current setup is greater than that of~\cite{super2}, it is smaller than~\cite{lensfree3} and thus is a fundamental limitation of the current setup.

In this study, we introduce a Bayesian-based method for performing aberration correction and numerical diffraction that accounts for interferometric aberrations, acquisition noise, and reconstruction artifacts to improve the effective NA and SNR of the reconstructed image.  The ability to improve the effective NA and SNR of lensfree on-chip microscopy images using the proposed method so that the morphological characteristics of biological specimens can be better visualized and studied can further enhance the use and benefits of lensfree on-chip microscopy.  Furthermore, increasing lensfree on-chip microscopy image quality using the proposed method could lead to improvements in tasks used for data analysis such as cell counting~\cite{cellcount}.

A.W. conceived and designed the method.  A.W. and X.W. worked on formulation and derivation of method solution.  C.J. performed the sample preparation. F.K. designed the lensfree on-chip microscopy system.  F.K. and C.J. performed the data collection.  A.W. performed the data processing.  A.W., F.K., and C.J. performed the data analysis. All authors contributed to the writing and editing of the paper. This work was supported by the Natural Sciences and Engineering Research Council of Canada, Canada Research Chairs Program, and the Ontario Ministry of Research and Innovation. The authors thank Dr. Monica Emelko at the University of Waterloo for the gracious support on culture supply and laboratory facilities.

\pagebreak

\end{document}